\documentclass[preprint,showpacs,preprintnumbers,amsmath,amssymb,floats]{revtex4}
\usepackage{blindtext}
\usepackage{graphicx}
\usepackage[english]{babel}
\usepackage{amsmath}
\usepackage{amssymb}
\usepackage{color}

\newcommand{\be}{\begin{eqnarray}}
\newcommand{\ee}{\end{eqnarray}}
\usepackage[toc,page]{appendix}

\begin{document}

\title{Universality and Crossovers for Quantum-Criticality in 2d metals}
\author{Chandra M. Varma}
\thanks{Visiting Scholar}
\affiliation{Department of Physics, University of California, Berkeley, CA. 94720}
\date{\today}
\begin{abstract} 
\noindent
 A simple generalization of the theory of crossovers in classical-criticality to quantum-criticality gives that, a Heisenberg model with a small anisotropy favoring planar order has a cross-over  towards the fixed point of the xy  model in the temperature direction which is very rapid compared to those in the orthogonal directions, if the temporal correlation length is much larger than the spatial correlation length, i.e. for a large dynamic exponent $z$.  At the other end of the flow, the stability of the fixed point of the quantum xy model coupled to fermions
 is exponentially enhanced in the temperature direction. This is used to explain why the quantum-critical fluctuations of all measured 2d anti-ferromagnetic compounds -  cuprates, heavy-fermion and Fe-based metals shows the characteristic fluctuations of the quantum xy model, and have the same anomalous transport and thermodynamic properties as the cuprates and twisted WSe$_2$ and Graphene. We segue briefly to the range of extended quantum-criticality due to disorder by generalizing the Harris criteria as well, using the properties of the quantum xy model. The observed $T \ln T$ specific heat at criticality is derived quite simply using the same methods which derive the cross-overs.
 
This paper is written for the commemoration volume for Jan Zaanen whom I knew  very well, starting from his days as a post-doc at Bell labs to his career  as a distinguished Professor at Leiden. 
\end{abstract}
\maketitle

\section{Introduction}
Twisted WSe$_2$ \cite{Pasupathy2021, Wang:2020, Mak_WSE2_2022} and Graphene \cite{Cao:2020, Efetov2022}, several heavy-fermion antiferromagnets \cite{LohneysenRMP2017, Paschen2010, Varma_IOPrev2016}, Fe-based anti-ferromagnets/superconductors \cite{Shibauchi_Rev:2014}, and the cuprates have the same anomalous properties in the normal state in their quantum-critical region which is also intimately related  to their superconductivity. This, despite the fact that the microscopic physics in these  systems  is quite dissimilar.  We know  that the properties do not follow from the critical fluctuations which are  the quantum extensions of the Ginzburg-Landau-Wilson paradigm.  Essentially all the important experimental facts follow from the fluctuations derived for  the quantum xy model coupled to fermions in which topological excitations in space and time determine the fluctuation. The final answers  are closely related to the phenomenological Marginal Fermi-liquid (MFL) spectrum, proposed earlier \cite{CMV-MFL, Kotliar-epl1991} to understand the quantum-critical properties of the cuprates. This is not a surprise for twisted WSe$_2$ where the microscopic physics imposes the xy model \cite{Chubukov_V2025} or twisted graphene if, as has been suggested \cite{Zaletel2020}, the critical order parameter for it, like cuprates, has a form of loop-current order \cite{Zhu-A-V2013, CMV_Graphene2023}, which also map to the xy model. But heavy-fermion and Fe compounds are usually discussed as  Heisenberg anti-ferromagnets coupled to fermions. Why do they belong to the same universality class?

 The MFL spectrum fit many experiments, predicted various other properties,  but had left open many important questions - it was an answer looking for a theory, as phenomenology usually is.
Even the fact that cuprates is a problem of quantum-criticality was not obvious, as it is in the  anti-ferromagnets,  because the quantum-critical order
parameter in cuprates was not  previously known or obvious from the very many experiments that had already been done, and had to be invented.  The  fact is that even though the order parameter and the microscopic physics are obviously quite different, the 
experimental properties in
the quantum-critical region, such as resistivity and specific heat, and the symmetry of superconductivity appear identical  for all these systems, when scaled by a microscopic parameter.  As will be reviewed below, the critical fluctuations measured by neutron scattering in the few quasi-two-dimensional anti-ferromagnets which are available are consistent with what were originally derived for the
2d-quantum xy model coupled to fermions to address the quantum-criticality of the cuprates.  Understanding the applicability to  the heavy-fermion and Fe-based antiferromagnet/superconductors requires  understanding  the cross-over to such a model over the entire observable range even though the anisotropy favoring it is small. 

A primary purpose of this paper is first to collect evidence that the quantum-critical fluctuations of the 2d metallic anti-ferromagnets are given by the solution of the 2d-quantum xy model coupled to fermions. To that end, a brief summary of the relevant results of that solution are first given. This is followed by the collection of fluctuation spectra directly measured. The first results reviewed are the most surprising and paradoxical, those in cuprate anti-ferromagnets near their quantum-critical point, as distinct from the other quantum critical point of the cuprates around which superconductivity is prominent. I will not discuss this other qcp of the cuprates, which has been adequately reviewed \cite{CMV-RMP2020}. This is followed by the measured spectra in the heavy-fermion anti-ferromagnet CeCu$_{6-x}$Au$_x$  and in an Fe based compound. There are several other metallic anti-ferromagnets in which the transport and thermodynamic properties are nearly identical but the fluctuation spectra has not been measured. The direct measurement of the critical spectra over its range of  measurements is hard and until recently done only by inelastic neutron scattering. Polarized inelastic x-ray scattering with sufficient energy and momentum resolution is coming of age and it might help in the future. 

The available data poses the  question: why the xy model for the anti-ferromagnets (for almost the whole observed range), which have Heisenberg coupling with perhaps only a small anisotropy towards planar order. We answer this be considering  the cross-over in criticality for the quantum model which has the remarkable property that the temporal correlation length are exponentially larger than the spatial correlation length.  

 I will also briefly mention the properties of  the universality of twisted WSe$_2$ and Graphene, where no direct measurements of the fluctuation spectra are available. 
 
  I will also discuss the effects of disorder on such critical fluctuations and why they appear to give an extended range of criticality for the quantum xy model. 

As a little diversion, it is shown how  the specific heat $~ T \ln (\frac{\omega_c}{T})$, which has hitherto been derived from the MFL self-energy of fermions \cite{CMV-MFL}, follows directly from the correlation function of fluctuations.

\section{Summary of results for the critical correlations of the quantum xy model coupled to fermions}

\begin{figure}
 \begin{center}
 \includegraphics[width= 0.8\columnwidth]{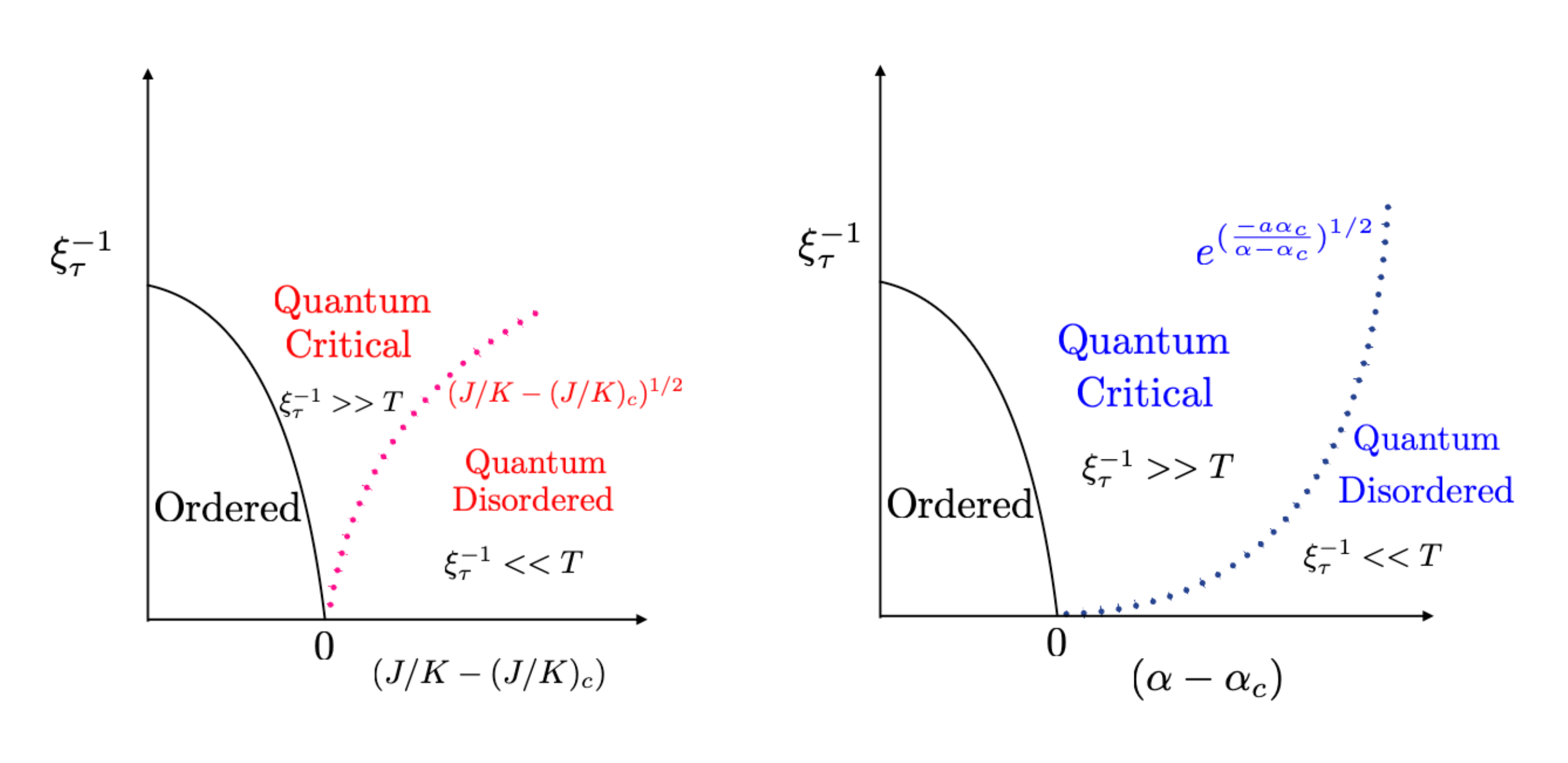}
 \end{center}
\caption{The quantum-critical, the ordered region and the quantum disordered region of the  quantum-xy model coupled to fermions, calculated by quantum Monte-carlo.
The phase diagram illustrates the temporal correlation length $\xi_{\tau}$ in various regions. The regions can be tuned by changing $J/K$, the ratio of the kinetic energy to the potential energy of the xy model or by changing the parameter $\alpha$ which depends on the residual resistivity of the metallic state. The variations of $\xi_{\tau}$ in these two directions are quite different, as shown in the left part and the right part of the figure. }
 \label{QC}
\end{figure}
\subsection{The Model}

The quantum-xy model coupled to fermions has three  parameters, $J$ - the Josephson coupling between the rotors, $K$ - the kinetic energy parameter for the rotors and $\alpha$ - the coupling constant of the fluctuations to the fermions. All are dimensionless parameters, $J$ and $K$ with respect to $\omega_c$, the ultra-violet cut-off in the problem and $\alpha$ which is proportional to the residual conductivity, determined by impurity density, measured in units of the quantum of conductivity. The phase diagram  of the model is illustrated in Fig. (\ref{QC}) focussing on quantities which are important for the discussion of cross-overs and of disorder given below.

The order parameter is the relative angle of the rotors $e^{i \theta}$. With only $J \ne 0$, the problem is classical and solved by Kosterlitz-Thouless \cite{KT1973, JKKN1977} as a topological transition driven by vortices, which  itself is not covered by the Ginzburg-Landau-Wilson paradigm. 

The action of the quantum model is
 \be
 S = S_{qxy} + S_{cf}.
 \ee
 $S_{qxy}$ is the action of  rotors of fixed length and angle $\theta({\bf x}, \tau)$ at a point ${\bf x}$ and imaginary time $\tau$, which is periodic in the inverse temperature $(0, \beta)$,  and is given by 
\begin{eqnarray}
\label{modelxy}
S_{qxy} = J \sum_{\langle {\bf x, x}' \rangle} \int_0^{\beta} d \tau \cos(\theta_{{\bf x}, \tau} - \theta_{{\bf x}', \tau})
 + K  \sum_{{\bf x}} \int_0^\beta d \tau \left( \frac{d \theta_{{\bf x}}}{d\tau}\right)^2 .
\label{eq:model}
\end{eqnarray} 
$S_{cf}$ gives the action for the coupling to fermions, which has the Caldeira-Leggett form \cite{CaldeiraLeggett} type, although the physical basis is quite different.  It is given in momentum and (Matsubara) frequency space by
\be
\label{scfqw}
S_{c-f} = \sum_{{\bf q}, \omega_n} \frac{\alpha}{4\pi^2} |\omega_n| q^2 |\theta({\bf q}, \omega_n)|^2.
\ee 
Here $\alpha$ is the  conductivity of the fermions in the limit $q \to 0, \omega \to 0, T \to 0$ made dimensionless in terms of $e^2/h$. $S_{c-f}$ looks complicated when written in terms of imaginary time, and need not be given here.

Extensive Monte-carlo results are available for the model \cite{ZhuChenCMV2015, ZhuHouV}. After a Villain transformation and integration over the spin-wave type fluctuations, the model is expressed \cite{Aji-V-qcf1, Aji-V-qcf3} in terms of interactions of two orthogonal topological excitations, vortices interacting purely in space and less familiar excitations, warps, which interact purely in time. Warps are local phase slips in units of time. As is shown \cite{Aji-V-qcf2}, they  may also be looked on as the annihilation or creation in time of the divergence of vortex current due to coupling to fermion-currents given by $S_{cf}$. The re-expression of the action in terms of vortices and warps allows a renormalization group calculation \cite{Hou2016}, such as by Kosterlitz \cite{Kosterlitz1974} for the classical problem of vortices alone. The answers agree with the Monte-carlo results although the
latter provide more details.

\subsection{Correlation functions at criticality}

The model has several phases \cite{ZhuChenCMV2015}.     We are interested here only in the fluctuation spectra near the quantum-disordered to ordered in space and time transition on which Fig. (\ref{QC}) concentrates.
This phase transition point at $T=0$ lies on a surface $F(\alpha_c, J_c, K_c) = 0$.
The surface is given in Fig. (8) of Ref. \cite{ZhuHouV}. 
Only the departure of the correlation lengths as a function of 
 $(\alpha -\alpha_c)/\alpha_c$ for a fixed  $(J/K)_c$ and $(J/K - (J/K)_c)$
 for a fixed $\alpha_c$ have been calculated.
The retarded correlation functions of the order parameter are calculated as \cite{ZhuChenCMV2015, ZhuHouV, Hou2016}
\be
\label{corfn}
 \chi({\bf r-r'}, \tau-\tau') &\equiv & <e^{- i\theta({\bf r'}, \tau')} e^{i\theta({\bf r}, \tau)}>, \\  \nonumber
 &=  & \chi_0 \frac{\tau_c}{\tau - \tau'} e^{-(\frac{\tau - \tau'}{\xi_{\tau}})^{1/2}}\ln \frac{a}{|{\bf r-r'}|} e^{-(\frac{|{\bf r - r'}|}{\xi_{r}})}.
\ee
$\tau$'s are imaginary times periodic in $2\pi/T$.

The most important feature of the order parameter correlations in Eq. (\ref{corfn}) is that they are products of a function of time and of space, quite unlike correlation functions for any problem, classical or quantum derived before. This is directly tested in experiments where correlation functions have  been measured, as shown below; it is also essential in getting the experimental properties like resistivity, and specific heat near the quantum-critical point. Also essential is the fact that the spatial correlation length is logarithm of the temporal correlation length \cite{ZhuChenCMV2015, ZhuHouV, Hou2016}:
\be
\label{spacetime}
 \frac{\xi_{r}}{a} = \ln \frac{\xi_{\tau}}{\tau_c}.
\ee
The calculated dependence of $\xi_{\tau}$  for the two different directions of approaching the critical point is different:
\be
\xi_\tau \propto (J/K- (J/K)_c)^{-1/2}
\ee
In this case the spatial correlation length grows only logarithmically with $(J/K- (J/K)_c)$; so the exponent $\nu \approx 0$.
The dependence in the direction $(\alpha - \alpha_c)$ is as
\be
\xi_\tau \propto e^{a \alpha_c/(\alpha - \alpha_c)^{1/2}}.
\ee
$a$ is a constant of $O(1)$. In that case, the exponent $\nu = 1/2$. The ordered, quantum-disordered and the quantum-critical region of the model and the variations of the temporal correlation length are sketched in Fig. (\ref{QC}).

Below, we shall denote both $J/K$ and $\alpha$ as $p$, the parameter which goes to $p_c$ at their critical values. 

At criticality the temporal part of the correlation function is $\propto 1/\tau$, the imaginary part of which on analytic continuation to real frequencies is $\tanh(\omega/2T)$ (below the ultra-violet cut-off $\omega_c$).
 This is precisely the marginal Fermi-liquid spectrum suggested much earlier. To get both properties like the long-wave-length fluctuations as well as the peculiar nuclear relaxation rate and the resistivity, a nearly space-independent fluctuation spectrum was suggested. That is hardly possible  at any form of criticality (except for some quantum impurities tuned to special points); the relation (\ref{spacetime}) gives a correlation length of the order of a lattice constant except exponentially close to criticality, and in effect serves the same purpose.

Eq. (\ref{corfn}) cannot be analytically continued to real frequencies except numerically. An approximate fit to the numerical results \cite{ZhuChenCMV2015} gives
\be
\label{corfn-1}
Im \chi({\bf q}, \omega) \approx \chi_0 \big(\ln \big|\frac{\omega_c}{max(\omega, \pi T, \xi_{\tau}^{-1})}\big| - i \tanh\frac{\omega}{\sqrt{(2T)^2 + \xi_{\tau}^{-2}}}\big) \frac{1}{q^2 + \xi_r^{-2}}.
\ee
For AFM critical fluctuations $q$ is measured from the putative Bragg-vectors of the transition.

A crucial testable prediction of (\ref{corfn-1}) is that the spatial correlation length $\xi_r$ near the transition is independent of frequency and temperature.
Another is the $\omega/T$ scaling form of the fluctuations and its specific form.

\subsection{Specific heat from correlation lengths}
 On length scales $\xi_r, \xi_{\tau}$, the scaling for the action or the normalized free-energy density, is a simple generalization to quantum-critical phenomena of the form for classical dynamical critical phenomena, see for example \cite{Goldenfeld}. (But we know for our model much more than just the scaling relations.) We are dealing with nearly degenerate fermions so that the density of excitations is $\propto N(0) T$, where $N(0)$ is the density of states of fermions at the chemical potential. Therefore, the free-energy has the scaling form
\be
\label{SpTA}
{\cal F}(p,T) \propto (N(0) T) \xi_r^{-d}\xi_{\tau}^{-1} f(\delta p_i \xi_r, T \xi_{\tau}).
\ee
Neither the factor $(N(0) T)$, nor $\xi_{\tau}^{-1} \propto T$ need be considered for properties near classical phase transitions, but they are crucial for  the thermodynamic properties at quantum phase transitions. 
One would normally have $\xi_{\tau} \propto \xi_r^z$. In our case $\xi_{\tau} \propto e^{\xi_r}$.
Let us be at $\delta p_i =0$ at a finite $T$, so that
\be
{\cal F}(p_c,T)  \propto   (N(0) T) \xi_r^{-d} e^{-\xi_r} f_T (T e^{\xi_r})
\ee
This gives us that the Free-energy at critical $p$ is
\be
{\cal F}(p_c,T) \propto T^2  \ln (\omega_c/T).
\ee
The specific heat is therefore $\propto T \ln (\omega_c/T)$, as observed, where $\omega_c$ is the ultraviolet frequency cut-off  of the fluctuations. This result was first derived by considering the self-energy of the fermions in MFL. and is found in all the diverse compounds discussed here where it has been measured. That it can be derived  as above should lend confidence to the results below in this paper, derived with the same simple technique.

\section{Universality in measured properties}

\subsection{Twisted-WSe$_2$ and Moir`e Graphene at their quantum-critical points}

No measurements of the critical fluctuations appear to be possible in these compounds. No direct measurements of the order parameter are available either. The resistivity as a function of temperature as well as a function of magnetic field has the same behavior as in all the other materials discussed here, see for example, for Graphene, Refs. \cite{Cao:2020, Efetov2022}, and for WSe$_2$, Refs. \cite{Pasupathy2021, Wang:2020, Mak_WSE2_2022}. I discuss now why these materials have xy type order to explain why they may be in the same universality class as the rest. 

In Twisted -WSe$_2$,  large spin-orbit coupling \cite{Devakul:2021} pins the z-direction of the spins so that they are not involved in the dynamics. So, as discussed, the model falls automatically in to the quantum-xy criticality class \cite{Chubukov_V2025}.

One of the proposed orders for twisted-Moi're graphene, called IVC (Inter-valley Coherent) order \cite{Zaletel2020} has at the C-C scale a loop-current order \cite{Zhu-A-V2013} which maps to the xy model \cite{CMV_Graphene2023}.  The linear temperature dependence and field dependence of the resistivity has also been quantitatively analyzed \cite{Varma_RH2022}. An alternate order is the Kekule order which is a real bond-order. It belongs to the Ising universality class and is unlikely to have a theory giving such transport properties. 

\subsection{Cuprates near their {\it AFM} quantum-critcality}

\begin{figure}
 \begin{center}
 \includegraphics[width= 0.8\columnwidth]{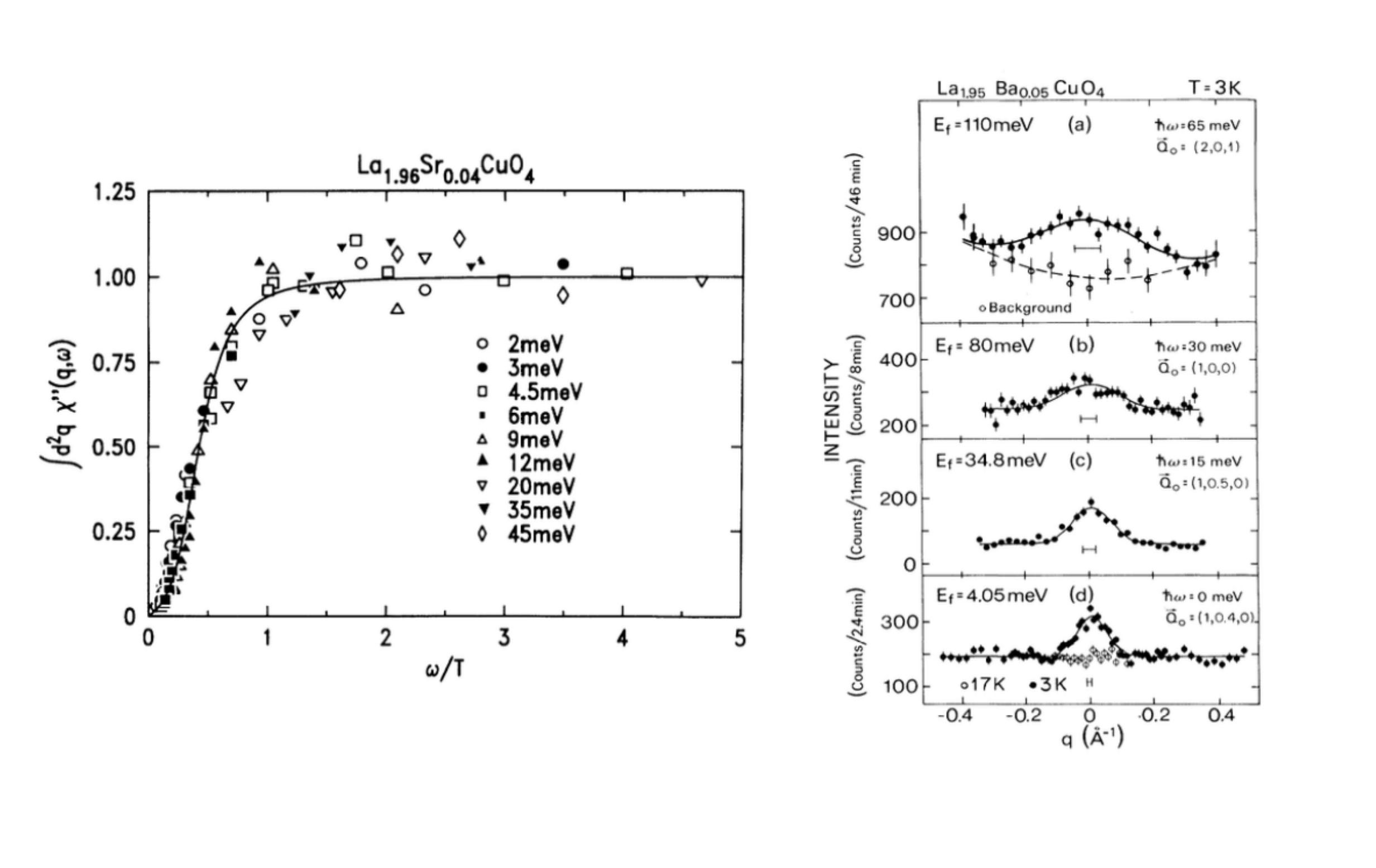}
 \end{center}
\caption{Left: The frequency and temperature dependence of the integrated over momentum fluctuation spectra in a lightly doped  cuprate compound near its antiferromagnetic criticality, taken from Ref. \cite{KeimerPRL1991}.
Right: Data showing that the momentum width of the critical fluctuations is nearly independent of frequency, taken from Ref. \cite{HaydenPRL1991}.}
 \label{lsco}
\end{figure}

The fluctuations of {\it anti-ferromagnetic insulating} Cuprates at half-filling  have been very well analyzed \cite{CHN1989} as the fluctuations of the Heisenberg anti-ferromagnetic order which would set up at $T=0$.
However, it orders (most likely due to the small three-dimensional coupling) at much higher temperatures. The ordering is always in-plane order in one of four equivalent $(\pi,\pi)$ directions which speaks for an anisotropy whose cause is not much investigated. Upon introducing carriers, the transition temperature rapidly declines and $\to 0$ for a doping of about 4\% per Cu-atom. (The antiferromagnetic correlation length at dopings of 15-20\% give correlation lengths of O(1) lattice constant, i.e. of O(($\pi k_F)^{-1}$), characteristic of non-interacting fermions.) Extensive inelastic neutron scattering for  La$_2$CuO$_4$ doped with Sr \cite{KeimerPRL1991} and with Ba \cite{HaydenPRL1991} and for YBa$_2$Cu$_{3}$O$_{6 +\delta}$, also close to the AFM transition are available. They all give similar results. Some of the results are shown in Fig. (\ref{lsco}). Both Refs. \cite{KeimerPRL1991, HaydenPRL1991} find that the correlation length of the AFM transition is independent of frequency and temperature of measurement but changes with doping near the quantum-critical point. This is in accord with the unusual result given above. To increase the signal/noise ratio, Ref.\cite{KeimerPRL1991} presents the frequency and temperature dependence integrated over momentum in range about the maximum value of the absorptive susceptibility, shown in left of Fig. (\ref{lsco}).  Clear $\omega/T$ scaling and close correspondence with the function (\ref{corfn-1}) is found. On the right of Fig. (\ref{lsco}), the spatial correlation length is shown to be nearly independent of frequency. Thus the critical fluctuations are of a product form in momenta and frequency and this together with other details shows that the AFM quantum-criticality belongs to the universality class discovered for the quantum-xy model coupled to fermions. 

At the dopings near the AFM qcp, the cuprates appear to have significant magnetic disorder and show signatures of a spin-glass type phase at low temperatures which persists to near optimal doping. The resistivity is linear in temperature only above about 50 K and increasing as temperature is decreased below it to that of an insulator, perhaps due to excessive disorder. There is no superconductivity. I think it would be interesting to suppress the AFM phase to its criticality in the undoped compound, perhaps by applying pressure, so that there is much less disorder and look at it for its normal state resistivity and possible high temperature superconductivity because the upper cut-off of the AFM critical fluctuations is expected to be about 2000 K.

A few relevant remarks about cuprates near optimal doping (15-20\% doping), to explain the absence of experiments observing such fluctuations by neutron scattering. Loop-current order was proposed as the critical order parameter because in the absence of spin-order or criticality,  fluctuations with magnetic component due to some other reason were required to get the NMR anomalies as well as all the others. Orbital current fluctuations were then a natural candidate. Such current orders require significant interactions between charge fluctuations on Cu and its neighboring O, which were an important part of the model favored \cite{cmv_lesHouches}.
The orbital order produces moments only of about 0.1 $\mu_B$ per unit-cell \cite{Bourges-rev}, about a factor of 5 smaller than the localized spin-moment at half-filling. Quantum-critical fluctuations of such moments distributed over the same energy scale as the spin-fluctuations of the insulator would then have an intensity 25 times smaller and not possible to detect. Evidence for the MFL fluctuations at long wave-lengths was found in Raman scattering \cite{Klein_1991}, and very impressively more recently over the entire momentum range in measurements of density-density correlations \cite{Abbamonte2018}. It should be noted however that such fluctuations, even though, they are not the fluctuations of the critical order parameter, acquire characteristics of the critical fluctuations at momenta other than those close to $0$ \cite{Shekhter2009, Varma_2017}.  The transport, thermodynamics and other properties for the cuprates have been discussed fully in Ref. \cite{CMV-RMP2020} and are quantitatively consistent with the theoretical results following from above.

\subsection{Heavy-Fermions AFM at their quantum-criticality}

\begin{figure}
 \begin{center}
 \includegraphics[width= 0.8\columnwidth]{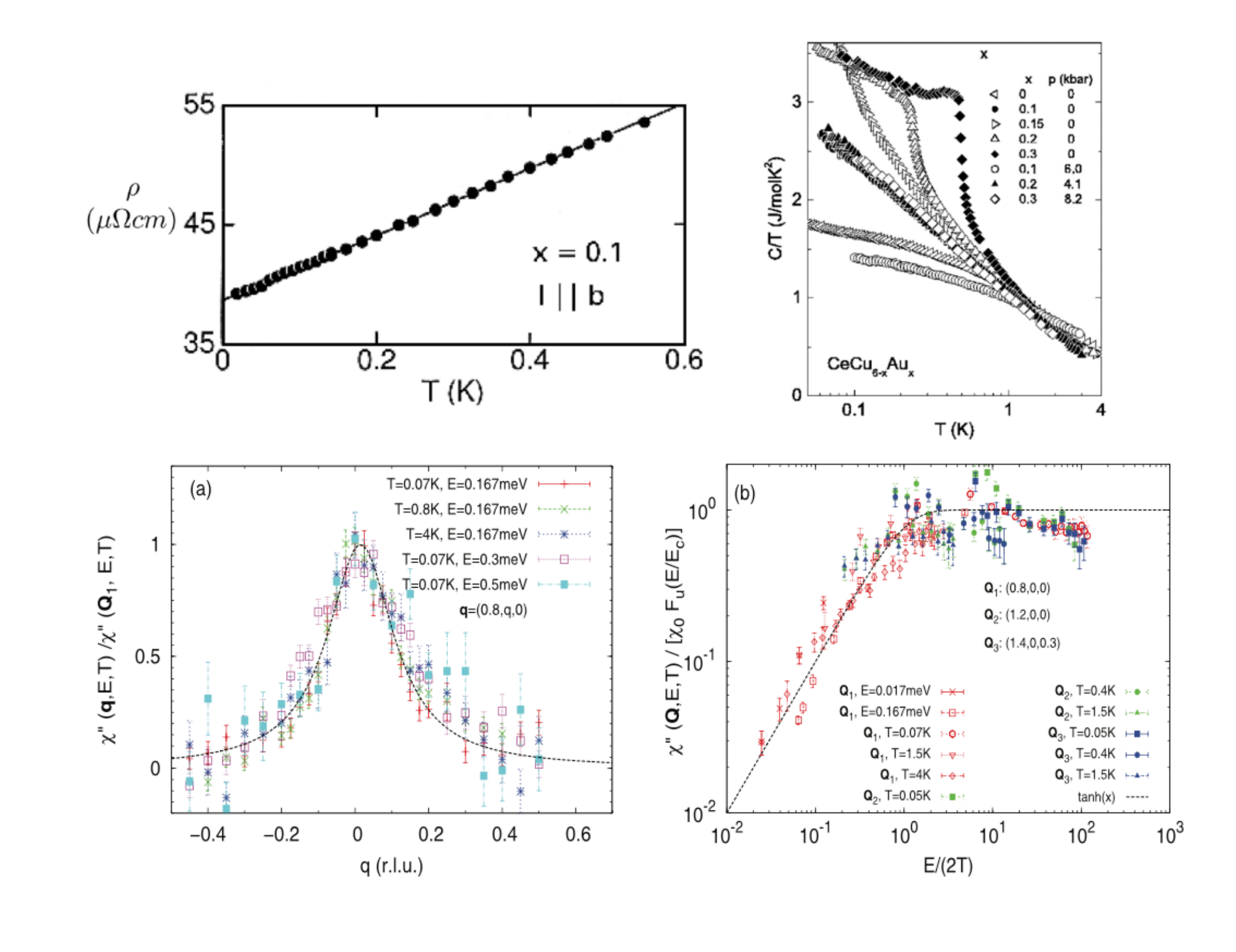}
 \end{center}
\caption{Top left: Resistivity of CeCu$_{0.9}$Au$_{0.1}$ and Right: Specific heat at various dopings showing the critical specific heat, the Fermi-liquid specific heat and the specific heat below the finite temperature AFM transition - data taken from \cite{LohneysenRMP2017}. Bottom left: Momentum dependence of critical fluctuations at various frequencies and temperatures, and right: Frequency and temperature dependence of the critical fluctuations - original data in \cite{Schroder1, Schroder2}, reanalyzed in 
\cite{SchroderZhuV2015}. The theoretical fit  on the right bottom uses the physical ultra-violet cut-off  because the measurements scan across that scale which is of the order given by the measured $T \ln (E_c/T)$ specific heat on the top right figure.}
 \label{cecu6}
\end{figure}

CeCu$_6$ is a heavy-fermion AFM which under pressure or substitution of some Cu with Au has an AFM qcp near which the resistivity is linear in T and specific heat varying as $T \ln (\omega_c/T)$ reproduced in Fig. (\ref{cecu6}) at the top. $\omega_c$ about 500  smaller than in cuprates, behooving a heavy-fermion compound whose characteristic Fermi-liquid energy scale - the Kondo temperature is only about 10 K. This small energy scale makes the magnetic fluctuations easy to measure and indeed more results are available for AFM quantum-criticality than almost any other compound. With collaboration of a co-author of the experimental paper, these results were drawn as shown in  the bottom of  Fig. (\ref{cecu6}) \cite{SchroderZhuV2015}.  The left part of the figure presents the momentum dependence of the critical fluctuations around the AFM vector ${\bf Q}_0$ for various $\omega$ and $T$ distributed over more than an order of magnitude. The correlation function in momentum is, well within the experimental noise,
independent of $\omega$ and $T$ as in the theoretical results for the xy model. The right part of the figure presents the frequency and temperature dependence normalized to its peak value as a function of ${\bf |q-Q_0|}$. This analysis again verifies the separation of the fluctuations as a product in momenta and in (frequency/temperature), a unique feature of the solution summarized above and indeed their detailed form. The measured specific heat and resistivity follow from such fluctuations. 

It is important in analyzing these measurements to use the upper cut-off $\omega_c$ because the measurements span through that scale. Indeed, an ultra-violet cut-off scale (related to the bare parameters of the model) must be an essential part of the any quantum-critical theory.
This was not kept in mind in a fit to the results given earlier in which both the infra-red and ultra-violet cut-offs are both provided by the temperature of measurements \cite{Schroder1}.
This can never be true in any physical theory. Such fits also find an exponent $\approx 0.75$ to the frequency dependence and therefore does not get the temperature dependence of the resistivity.

The correct theory of a single Kondo impurity in conduction electrons near their AFM quantum-criticality \cite{SiNature2001},  whose results were  reproduced by a different method in  \cite{ Maebashi2}, (called "Kondo deconstruction") has been sometimes used to analyze AFM quantum-critical points in heavy-fermions. It does not get the linear in T resistivity, but even more to the point, it is  flawed because
if two interacting impurities or three or four are used in the theory, the results change dramatically. Using the criticality of the two Kondo impurity problem  in a mean-field calculation \cite{Delft_2024} is only  better because the solution of the two-impurity problem \cite{Jones_V_Wilkins} at least yields (an unstable) fixed point with MFL fluctuations, but it needs unphysical tuning. It cannot be even an approximate solution to the physical problem at hand; results in the theory flow away on both sides to Fermi-liquids at low temperatures as in the solution of the two Kondo impurity problem; one never has an AFM. 

Many other heavy-fermion compounds near quantum-criticality have similar thermodynamics and transport as discussed above; special mention must be made of YbRh$_2$Si$_2$ and its alloys  \cite{Paschen2010} which very extensive such measurements and magneto-oscillation measurements are available Unfortunately there are no direct measurements of the fluctuation spectra.

\subsection{Fe-based AFMs at their quantum-criticality}

\begin{figure}
 \begin{center}
 \includegraphics[width= 0.8\columnwidth]{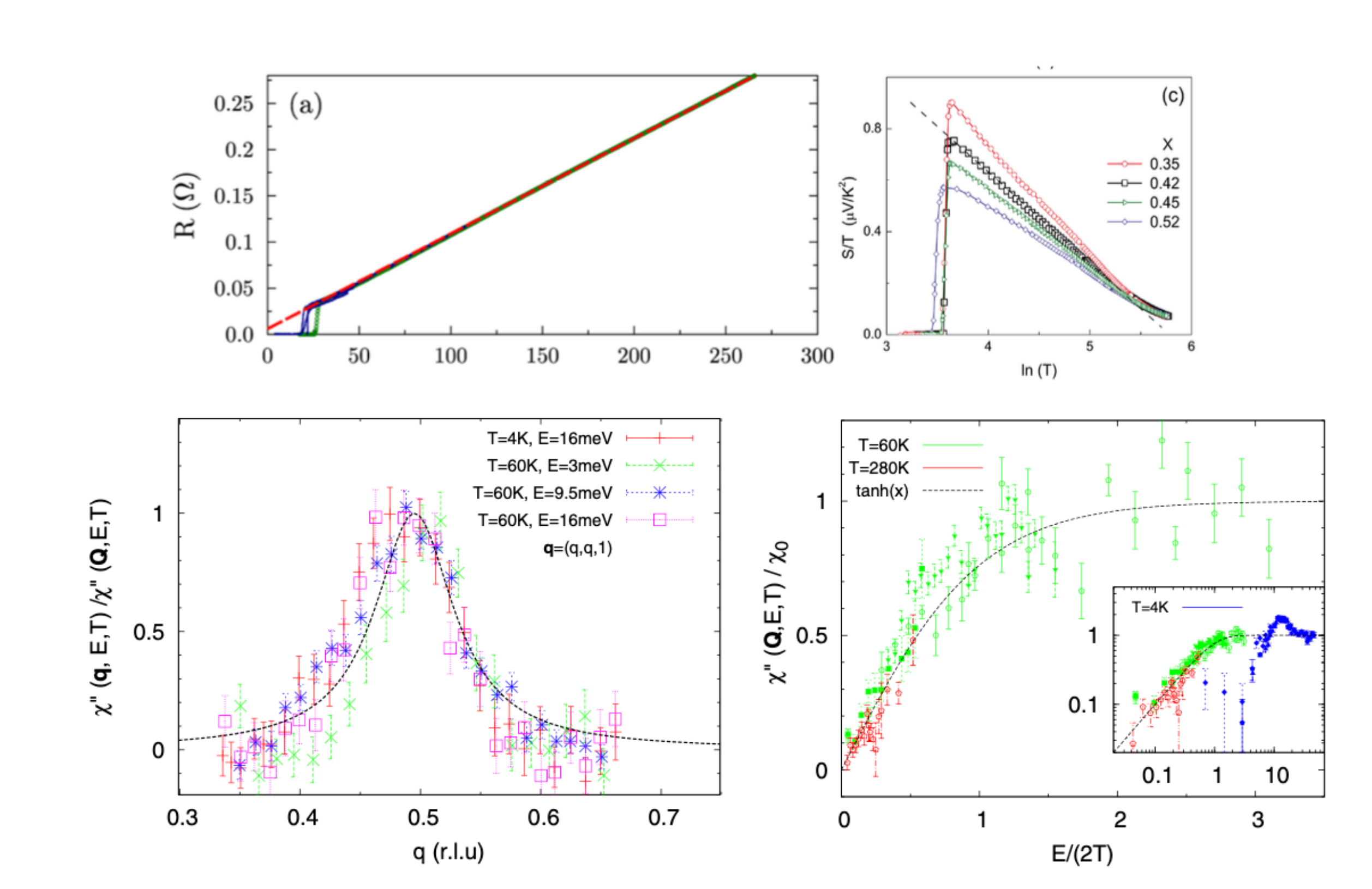}
 \end{center}
\caption{Top: left, the resistance vs. temperature at various magnetic fields  for the compound Ba(Fe$_{1-x}$Co${_x}$As$_2$) for $x =0.41$ close to its critical composition. right: the Thermopower (proportional to electronic contribution to the entropy) at various dopings in the same compound - both taken from \cite{BaAsFe}. Bottom: left, Momentum dependence of the critical fluctuations at various temperatures and frequencies specified and, at right, the frequency and temperature dependence, at a nearby composition, taken from Ref. \cite{Inosov2010}. The inset shows data at larger $E/2T$ in blue, which continues the normal state data in the superconducting state except for a bump at near twice the superconducting gap energy and suppresses it below the bump.}
 \label{Fecomp}
\end{figure}

Extensive enough measurements are available in an Fe-based AFM/superconductor near its qcp to be analyzed systematically.  The compound is BaFe$_{1-x}$Co$_{x}$As$_2$. The resistivity and the thermopower near criticality are shown in top of Fig. (\ref{Fecomp}) and the measurements of critical fluctuations by neutron scattering at the bottom in the same figure. The superconducting transition temperature of 25 K cuts off the low-frequency fluctuations, giving a peak in the frequency
dependence neat twice the gap at about 10 meV. The momentum dependence at temperature and frequency varying by an order of magnitude  shown in the left part of 
the figure,  the $\omega/2T$ dependence over a factor of about 30 is shown in the right part of the figure. The upper cut-off scale of the fluctuations is too large to be measured by neutron scattering. All that was said about the heavy-fermion compound in relation to its figures can again be said here.

\section{Crossovers for Quantum-critical points of AFMs due to anisotropy}

Let us consider a nearly isotropic Heisenberg model in a metal which has a quantum critical point as the  dimensionless parameter $p$ approaches the isotropic critical value $p_i$. The deviation from the Heisenberg model is in the form of  a single-ion or exchange anisotropy. In classical transitions, the flow-diagram is the same for either form of anisotropy \cite{Goldenfeld}.  Let us consider the free-energy density $F(p,T,A)$, as a function of the departure from quantum criticality and as a function of the dimensionless anisotropy parameter $A$ with xy coupling larger than the zz coupling. 
This choice is natural since the long-range order in all the metallic AFMs we consider in in the plane. $A <<1$ may be taken as the ratio of the anisotropic coupling energy to the isotropic coupling energy. Similarly $T$ is dimensionless, being the ratio of the physical temperature to the ultra-violet cut-off energy $\tau_c^{-1}$, which is also similar to the isotropic coupling energy. 


We now augment Eq. (\ref{SpTA}) with a function depending on the anisotropy with an exponent $y_{A_i}$. For classical phase transitions we know that $y_{A_i} > 0$  and there is no reason this should not be so for the quantum problems.
\be
\label{SpTA2}
S(p,T, A) = {\cal F}(p,T,A)   \propto (N(0) T) \xi_r^{-d}\xi_{\tau}^{-1} f(\delta p_i \xi_r^{y_{pi}}, T \xi_\tau^{y_{\tau i}}, A \xi_r^{y_{A_i}}).
\ee
 Also introduced are the dynamical critical exponent $z_i$ for the isotropic problem and the renormalization group eigenvalues $y_{pi} = 1/\nu_i$ and 
$y_{Ai}$ for the operators $p$ and $A$, respectively. 
For quantum criticality, we have to calculate the relevance of perturbations both as a function of $(p-p_c)$ and as a function of $T$. 
As explained in the previous section, $p$ may be either $\alpha$ or $J/K$. Different cross-overs flows are expected in the four different cases. 
We need the critical parameters for a Heisenberg model coupled to fermions, the product  $\nu_i y_{A_i}$ and the dynamic exponent $z_i$ to determine the flow. 

For classical transitions in models with n component spins and in (4-$\epsilon$) dimensions, the cross-over exponent  $\phi_i \equiv \nu_i y_{Ai}$, calculated to $O(\epsilon)$ is \cite{Fisher_RMP},
\be
\phi_i = 1 +\frac{n \epsilon}{2(n+8)}.
\ee
This helps in giving an indication of the speed and direction of the flow. 

Consider first the quantum-disordered region in which $\xi_{\tau} >> T$, in which we may put $T \xi_{\tau} = 0$  and examine the flow of anisotropy as $p \to p_c$. Under these conditions,
\be
\label{SpTA1}
S(p,T,A) = {\cal F}(p,T,A)/k_BT  \propto \xi_r^{-d}\xi_{\tau}^{-1} f(\delta p_i \xi_r^{y_{pi}}, 0 , A \xi_r^{y_{A_i}}) = \xi_r^{-d}\xi_{\tau}^{-1} f_p(\delta p_i \xi_r^{y_{pi}}, A \xi_r^{y_{A_i}})
\ee
Assuming scaling holds,
\be
\label{SpTA2}
S(p,T,A) \propto  f_p(\frac{A}{\delta p^{\nu_i y_{Ai}}}).
\ee
Since $\nu_i y_{Ai} > 0$, anisotropy grows as $\delta p_i$ decreases to approach the transition. This is completely akin to cross-over in classical problems in which $\delta p_i$ is the dimensionless deviation from the critical temperature \cite{Goldenfeld}.
 
In the quantum-critical region near $\delta p =0, T ~\xi_{\tau} \to 0$; assuming $z_i > 1, T \xi_{r}  << 1$ also. So we will put $\delta p~ \xi_r \to 0$ to get the flow of anisotropy in the $T$-direction given by:
\be
\label{F-T}
S(p,T,A) \propto f_T\big(\frac{A_i}{T^{(y_{A_i} /z_i)}}\big).
\ee
So as $T$ decreases, anisotropy is reduced faster than in the quantum-disordered region because $z_i > 1$. For Heisenberg model coupled to fermions,
$z_i = 3$ has been shown \cite{Hertz, Millis-qcf}.

The scaling about the isotropic point is towards the anisotropic critical point. One may rightly question the scaling ansatz when $z_i =3$. It is hoped that at least the direction of the flow is correctly indicated. Since we know the correlation functions near the fixed point towards which this flow is directed, the stability shown below may be regarded as on firmer ground.

Let us consider if the anisotropic critical point is stable by using the results for the correlation function calculated for the xy model. We must now derive how $\overline{A} < < 1$ flows, where $\overline{A}$ is the ratio of the zz-coupling to the xy coupling. Near the planar critical point $y_{\overline{Aa}} < 0$, because the anti-ferromagnetic order in all the known examples is planar. 
 
One finds by the same procedure as above that the irrelevance of  $\overline{A}$ around $\delta p_a =0$ is given in the quantum-disordered region by
\be
\label{psc}
 f_p \big(\frac{\overline{A}} {\delta p^{\nu_a |A_{a}|)}}\big).
  \ee
  So the anisotropy is irrelevant about the  xy point. ($\nu_a$ in this case is either $1/2$ or may be taken to be zero for logarithmic growth with $(p-p_c)$).
  In the quantum-critical region, the irrelevance is stronger,
  as $\overline{A}$ goes down as
  \be
  \label{tsc}
  f_T\big(\overline{A} Te^{- 1/(T |A_{a}|)}\big),
  \ee
  which together with (\ref{psc}) implies that if we consider a finite fixed $\delta p$, the anisotropy is also irrelevant for $T \to 0$.

Putting the scaling near the isotropic and anisotropic critical points together, the conclusion is that in the $p-$direction, the crossover is similar to the crossover in temperature of the classical problem. On the other hand, in the temperature direction, $z_a >> 1$ leads to correlation functions and therefore properties calculated from it, such as the temperature dependence of the resistivity and of the specific heat, which are those of the anisotropic critical point over a wide range. Near the anisotropic critical point, the dependence of the correlation lengths on $(p-p_c)$ are explicitly known; so we do not have to use $z >> 1$.


\section{Crossover due to Disorder for quantum critical points}

Suppose the quantum critical point $p_c$ is a function of disorder. The analogous classical problem is the problem in which the parameter determining the transition temperature varies locally, which was formulated by Harris \cite{Harris_1974}.  The correlation functions appear, as for the classical phase transitions, strongly dependent on the nature of disorder. Consider here only disorder with short-range correlations similar or smaller than $\xi_r$, with variations in $p_c({\bf r})$ which are Gaussian correlated. On a length scale $\xi_r$, the root mean-square fluctuation of $p_c$, $\Delta p_c = w_0 \xi_r^{-d/2}$ in the classical problem.  This does not change for quantum-critical phenomena because disorder remains perfectly correlated in time. Two questions need to be asked: what is the scaling of $\Delta p_c$, which gives the variance of the transition point at $T=0$ compared to $(p - p_c)$. For this, one must take $T \to 0$ first. Secondly how at a fixed $p$, not too far from $p_c$, $\Delta p_c$ scales with temperature. If it goes to $0$ faster than $T$, the critical properties extend over a range of $p$ around $p_c$.

This is the usual story. In our problem, the situation is different because the critical point is determined by $\alpha \to \alpha_c$ and $\alpha$ as summarized in
Sec. (II) depends on the residual resistivity, in other words on disorder. If disorder is increased (within some limits which have  not been determined), the critical point $\alpha_c$ shifts with a corresponding change in $(J/K)_c$. The important point to note is that the  phase diagram for criticality determined by disorder is the second of the two shown in Fig. (\ref{QC}), where the critical region extends as an essential singularity. The increase in disorder makes the ordered region grow so that the critical region shifts towards larger doping. The quantum-critical region (in which the resistivity is $\propto T$ and the specific heat is $\propto T \ln(\omega_c/T)$, then is  insensitive to disorder.  Obviously this cannot go on as disorder continues to increase, but no calculations have been done to find the limit of the behavior.

There is much confusion in the literature on the region of $T$ linear behavior of resistivity, some of it caused by improper use of power laws in resistivity across regions of cross-overs of the phase diagram and, much worse, fitting resistivity in a region in which superconducting transition temperature is reduced by applying a magnetic field without taking into account the temperature dependence of large magneto-resistance
of 2d superconductors towards the normal state. Nevertheless in the over-doped region of the cuprates and in some heavy-fermion compounds, there is considerable evidence \cite{Hussey_2013} of asymptotic linear in $T$ resistivity over an extended range and a slow cross-over to it  from a quadratic $T$ dependence. For a given impurity density, the quantum-critical region in Fig. (\ref{QC})-right hugs the $T=0$ line because of the essential singularity in the form of the cross-over to the quantum-disordered side. I think this is the true explanation of such "extended criticality".

However, it is amusing to consider the disorder problem for general quantum-critical points in which disorder is not a tuning parameter for criticality but $z$ may be large, (with always heeding the warnings from criticality priest-hood, but not regarding it as killing, that scaling may not  apply).

The action has the form
\be
\label{action2}
S(p,T,\Delta p_c) = {\cal F}(p,T,\Delta p_c)/k_BT  \propto \xi_r^{-(d+z)} f(\delta p \xi_r^{y_{p}}, T \xi_{\tau}, w_0 \xi_r^{-d/2}\delta p^{-\phi_x}).
\ee
Here $\phi_x$ is the cross-over exponent for $\Delta p_c$. $\phi_x =1$ because $\delta p_c$ has the same dimension as $p$. 

For $T \xi_{\tau}^z << 1$, it is replaced by 0. 
Then  (\ref{action2}) gives that disorder scales with $|p-p_c|$ as
\be
\label{disord1}
S(p, \Delta p_c) \propto {\cal F}_p(w_0 |p-p_c|^{(d/2)\nu_p -1})
\ee
If $|p-p_c|^{(d\nu_p/2 -1}$ grows as $|p-p_c| \to 0$, the pure fixed point is unstable or that the
disorder is relevant for 
\be
\label{rel-p}
d\nu_p/2 -1  < 0.
\ee
This is the same as the Harris criteria for the relevance of disorder in classical problems as $T \to T_c$ modified to a quantum transition at $p \to p_c$.

Let us look at the crossover due to disorder in the temperature direction. In other words, we now take the limit $p \to p_c$ first, so that the dependence on $(p-p_c)$ is replaced by 0. Now $\Delta p_c$ scales with respect to temperature. Since $T$ scales with $\ell_r^z \propto (\delta p)^{z\nu}$ relevance for disorder in the time-direction occurs for 
\be
\label{rel-T}
w_0 \Big(\frac{T}{\omega_c}\Big)^{(d\nu_p/2 -1)/(\nu_p z)} > T. 
\ee
Here $\omega_c$ is the ultra-violet cut-off of the critical fluctuations in the problem. Suppose disorder is marginally relevant in the $p-$ direction.  Then, comparing (\ref{rel-p}) with (\ref{rel-T}), the conclusion is that for large $z$, one must go to smaller $T \tau_c$ than in $|p-p_c|/p_c$ to see the cross-over in temperature (or frequency) to notice the effect of disorder.

I thank Erez Berg for sending me a draft of a paper dealing with 'extended quantum-criticality', which led me to formalize what I had been thinking about that problem.

\end{document}